\documentstyle[multicol,psfig,epsf,aps]{revtex}
\newcommand{\la}{\left\langle}
\newcommand{\ra}{\right\rangle}
\newcommand{\be}{\begin{equation}}
\newcommand{\ee}{\end{equation}}
\newcommand{\bea}{\begin{eqnarray}}
\newcommand{\eea}{\end{eqnarray}}
\newcommand{\ba}{\begin{array}}
\newcommand{\ea}{\end{array}}
\newcommand{\bi}{\begin{itemize}}
\newcommand{\ei}{\end{itemize}}

\newcommand{\piulbl}{$\Pi^{u<}_{b<}$}

\begin{document}
\draft
\title{On Generation of magnetic field in astrophysical bodies}
\author{Mahendra\ K.\ Verma}
\address{Department of Physics, Indian Institute of Technology,
Kanpur  --  208016, INDIA}
\date{Nov. 26, 2001}
\maketitle

{\bf The generation of magnetic field in astrophysical bodies, e.g.,
galaxies, stars, planets, is one of the outstanding theoretical
problems of physics and astrophysics.  The initial magnetic fields of
galaxies and stars are weak, and are amplified by the turbulent motion
of the plasma.  The generated field gets saturated due to nonlinear
interactions.  The above process is called ``dynamo'' action.
Qualitatively, the magnetic field is amplified by the stretching of
the field lines due to turbulent plasma motion.  A fraction of kinetic
energy of the plasma is spent in increasing the tension of the
magnetic field lines, which effectively enhances the magnetic field
strength.  Current dynamo theories of are of two types, kinematic and
dynamic.  In the kinematic theories, one studies the evolution of
magnetic field under a prescribed velocity field.  In kinematic
$\alpha$-dynamo, the averaged nonlinear term $\la {\bf u \times b}
\ra$ (${\bf u,b}$ are velocity and magnetic field fluctuations
respectively) is replaced by a constant $\alpha$ times mean magnetic
field ${\bf B_0}$.  This process, which is valid for small magnetic
field fluctuations, yields linear equations that can be solved for a
given boundary condition and external forcing fields$^{1-3}$.  In
dynamic theories$^{4-6}$, the modification of velocity field by the
magnetic field (back reaction) is taken into account.  Using a
different approach, here we compute energy transfer rates from
velocity field to magnetic field using field-theoretic method.  The
striking result of our field theoretic calculation is that there is a
large energy transfer rate from the large-scale velocity field to the
large-scale magnetic field.  We claim that the growth of large-scale
magnetic energy is primarily due to this transfer.  We reached the
above conclusion without any linear approximation like that in
$\alpha$-dynamo.  }

There is an exchange of energy between various Fourier modes because
of nonlinear interactions present in magnetohydrodynamics (MHD).
Since there are two vector fields ${\bf u}$ and ${\bf b}$ in MHD, the
energy can be transferred from ${\bf u}$ to ${\bf u}$, ${\bf u}$ to
${\bf b}$, and ${\bf b}$ to ${\bf b}$.  Energy from a parent mode
${\bf u(k)}$ or ${\bf b(k)}$ ({\bf k} represents the wavenumber of
Fourier mode) is transferred to two daughter modes with wavenumbers
${\bf p}$ and ${\bf k-p}$.  The allowed triads in MHD are $({\bf u(k),
u(p), u(k-p)})$ and $({\bf u(k), b(p), b(k-p)})$.  The net effects of
all the energy transfers are constant energy fluxes from large-scale
$u$ to small-scale $u$ ($\Pi^{u<}_{u>}$), large-scale $u$ to
small-scale $b$ ($\Pi^{u<}_{b>}$), large-scale $b$ to small-scale $u$
($\Pi^{b<}_{u>}$), large-scale $b$ to small-scale $b$
($\Pi^{b<}_{b>}$), and large-scale $u$ to large-scale $b$
($\Pi^{u<}_{b<}$).  The superscript and subscript of $\Pi$ refer to
the source and sink respectively.  All these energy fluxes are
illustrated in Fig.~1.  These energy fluxes are analogous to that of
Kolmogorov's flux in fluid turbulence.  Note that large-scale velocity
modes are forced, as shown in the figure.

Stani\u{s}i\'{c}$^7$, Dar et al.$^8$, and others have given formulas
for computation of the above fluxes.  However, Dar et al.'s formalism
is the most general, and they have numerically computed all the fluxes
of MHD using numerical data of direct numerical simulation$^8$.  In
the present paper, we compute the MHD fluxes in the inertial range
using field-theoretic method.  Our calculation is up to first order in
perturbation.

We give a brief outline of the theoretical calculation (
refer to Verma$^9$ for details).  We write down the evolution
equation for kinetic energy spectrum ($\la {\bf |u(k)|^2} \ra /2$) and
magnetic energy ($\la {\bf |B(k)|^2} \ra /2$).  We carry out our
analysis for three space dimensions, constant mass density, and zero
mean magnetic field.  We also assume that $\la {\bf u \cdot b} \ra
=0$, and that the large-scale velocity modes are forced.  There is a
net outflow/inflow of energy from a wavenumber sphere (say sphere of
radius $k_0$) as discussed in the earlier paragraph. These energy
fluxes can be easily calculated using the Fourier modes.  To
illustrate, the energy flux from the modes inside of the $u$-sphere of
radius $k_0$ to the modes outside of the $b$-sphere of radius $k_0$ is
given by
\be 
\Pi^{u<}_{b>}(k_0) =
\int_{k'>k_0} \frac{d {\bf k}}{(2 \pi)^3} \int_{p<k_0} 
\frac{d {\bf p}}{(2 \pi)^3} 
\la \Im ({\bf [k \cdot u(q)][b(k) \cdot u(p)]}) \ra
\label{eqn:flux}		
\ee
where $\Im$ stands for the imaginary part, and ${\bf k=p+q}$.  Clearly
the sources of energy for the $b<$ sphere are $\Pi^{u<}_{b<},
\Pi^{u>}_{b<}$, and $\Pi^{b>}_{b<}$.  If there is a net flux of energy
into large-scale magnetic energy, then magnetic energy at large-scale
will grow, or dynamo is active.

We calculate the energy fluxes perturbatively to first order.  We
assume homogeneity and isotropy for the flow.  In the correlation
functions we have included kinetic helicity ($H_K=\la {\bf u \cdot
\omega} \ra/2$, where ${\bf \omega}$ is the vorticity) and magnetic
helicity ($H_M=\la {\bf a \cdot b} \ra/2$, where ${\bf a}$ is the
vector potential), which are defined using
\bea
\la u_i ({\bf k}) u_j ({\bf k'}) \ra &  =  &
        \left[ P_{ij}({\bf k}) C^{uu} ({\bf k}) -
       i \epsilon_{ijl} k_l \frac{2 H_K(k)}{k^2}  \right] 
	\delta({\bf k+k'})  \label{eqn:uiuj} \\
\la b_i ({\bf k}) b_j ({\bf k'}) \ra &  =  &
         \left[ P_{ij}({\bf k)} C^{bb} ({\bf k}) -
           i \epsilon_{ijl} k_l 2 H_M(k) \right] \delta({\bf k+k'}) 
	 \label{eqn:bibj}
\eea
Note that helicities break mirror symmetry.

We focus on the fluxes in the inertial range.  Based on recent
theoretical$^{10-14}$ and
numerical$^{15,16}$ evidences, we take Kolmogorov's
spectrum for the correlation functions in the inertial range, i.e.,
\bea
C^{uu}({\bf k})  =  \frac{K^u}{4 \pi} \Pi^{2/3} k^{-11/3}; &\hspace{1cm} &
C^{bb}({\bf k})  = C^{uu}({\bf k})/r_A; \\
H_K({\bf k})     =  r_K k C^{uu}({\bf k}); & \hspace{1cm} &
H_M({\bf k})     =  r_M C^{bb}({\bf k})/k 
\eea
Since both magnetic and kinetic energy spectrum are Kolmogorov-like in
the inertial range, $r_A, r_K$ and $r_M$ can be treated as constants.
We also use turbulent or renormalized viscosity and resistivity in our
calculation.  These quantities have been recently derived by
Verma$^{13,17}$.  Using the steady-state condition, we also calculate
the energy supply from the large-scale velocity field to the
large-scale magnetic field $\Pi^{u<}_{b<} =
\Pi^{b<}_{b>}+\Pi^{b<}_{u>}$.  Since the inertial-range energy
spectrum is universal, the above energy flux is independent of the
details of large-scale forcing.

We calculate various fluxes in terms of $r_A, r_K, r_M$ (see Verma$^9$
for further details).  In Fig.~1 we have shown both nonhelical (solid
line) and helical (dashed line) contributions for the case ($r_A=5000,
r_K=0.1, r_M=-0.1$).  Large $r_A$ (here $r_A=5000$) corresponds to a
very weak magnetic field compared to the velocity field, similar to
the early phase of galactic and stellar evolution.  The choice of
$r_K=0.1$ and $r_M=-0.1$ is motivated by the fact that both kinetic
and magnetic helicities in most of the astrophysical plasmas are
relatively small.  The recent simulations of Brandenburg$^{18}$ as
well as the EDQNM calculation of Pouquet et al.$^4$ shows that the
magnetic helicity is negative for small wavenumbers.  Since the small
wavenumber contributions dominate the contributions from the large
wavenumbers, we have taken $r_M < 0$.

The flux ratios shown in Fig.~1 illustrate many important results.
They are: 
\begin{itemize}

\item There is a large energy flux from large-scale velocity field to
large-scale magnetic field ($\Pi^{u<}_{b<}$).  In addition, there are
two other fluxes, $\Pi^{b>}_{b<helical}+\Pi^{u>}_{b<helical}$, to the
large-scale magnetic field.  These fluxes are responsible for the
growth of large-scale magnetic field in the initial stage of
evolution.  Pouquet et al.$^4$, Pouquet and Patterson$^{19}$, 
Brandenburg$^{18}$, and many others
generally highlight $\Pi^{b>}_{b<helical}$ inverse transfer, and do
not consider \piulbl.  Recently Brandenburg$^{18,20}$
argues that large-scale magnetic energy is sustained by nonlocal
inverse cascade from the forcing scale directly to the largest scale
of the box.  He relates this effect to the $\alpha$-effect.  In this
paper and Verma$^9$ we have computed the relative
magnitudes of all three contributions for  generic parameters
discussed above, and find that all of them to be comparable, however,
$\Pi^{u<}_{b<}$ is somewhat higher.

\item As shown in the figure, the nonhelical component of magnetic
energy flux ($\Pi^{b<}_{b>}$) is forward, while the helical component
of magnetic energy flux is negative (inverse).  The overall
magnetic energy flux however is positive.

\end{itemize}

In our theoretical calculation we have assumed homogeneity and
isotropy of the flow, as well as ${\bf B_0=0}$.  These assumptions are
likely to hold in the early stages of the galactic evolution before
large structures appear.  To model the early evolution of galaxies, we
assume that the large-scales contains kinetic and magnetic energies.
During this unsteady period, magnetic energy($E^b$) will be amplified
by the nonhelical $\Pi^{u<}_{b<}$, and helical components
$\Pi^{b>}_{b<helical}$ and $\Pi^{u>}_{b<helical}$, i.e.,
\be 
\label{eqn:dynamo_Ebdot}
\frac{d E^b(t)}{dt} =  \Pi^{u<}_{b<}+ \Pi^{b>}_{b<helical} + 
			\Pi^{u>}_{b<helical}
\ee
We assume quasi-steady state for the galactic evolution.  Our
theoretical calculation performed for large $r_A$ shows that the
energy fluxes of the right-hand-side of Eq.~(\ref{eqn:dynamo_Ebdot}) is
proportional to $\Pi E^b/E^u$, where $E^u$ is the kinetic energy, and
$\Pi$ is the total energy flux or energy supply rate.  Using
Kolmogorov's spectrum, we obtain
\be
E^b(t) \approx E^b(0) \exp{\left(\frac{ \sqrt{E^u}}{L (K^u)^{3/2}} 
		t \right)}
\ee
where $L$ is the large-length of the system. Clearly, the magnetic
energy grows exponentially in the early periods, and the time-scale of
growth is of the order of $L/\sqrt{E^u}$, which is the eddy turnover
time.  Taking $L \approx 10^{17} km$ and $\sqrt{E^u} \approx 10
km/sec$, we obtain the growth time-scale to be $10^{16} sec$ or $3
\times 10^8$ years, which is in the expected range$^{21}$.
Hence, we are able to construct a nonlinear and dynamically consistent
galactic dynamo based on the energy fluxes.  In this model, magnetic
energy grows exponentially, and the growth time-scale is 
reasonable$^{21}$.

To recapitulate, we address the dynamo problem in the light of
turbulent energy flux.  Our approach is very different from the
$\alpha$-dynamo picture.  Our theoretical calculation, based on
perturbative field theory, shows large amount of energy transfer from
large-scale velocity field to large-scale magnetic field.  This
transfer occurs in both helical and nonhelical MHD.  We believe
magnetic energy growth is primarily due to this energy transfer.
Regarding magnetic energy flux, there is (a) nonhelical forward flux
from large scales to small scales, and (b) helical inverse transfer
from small scales to large scales.  The net magnetic energy transfer,
however, is positive.

We have constructed a model of galactic dynamo based on our energy flux
results.  We find that magnetic energy grows exponentially, and our
estimate of its growth time-scale is consistent with the current
observational estimates$^{21}$.


\acknowledgements 
The author thanks G. Dar, V. Eswaran, Krishna Kumar,
R. K. Varma, and D. Narsimhan for discussions.  He also thanks Agha
Afsar Ali for carefully reading the manuscript.

Correspondence and requests for material should be addressed
to M.~K. Verma  (email: mkv@iitk.ac.in).

\begin{figure}
\centerline{\psfig{figure=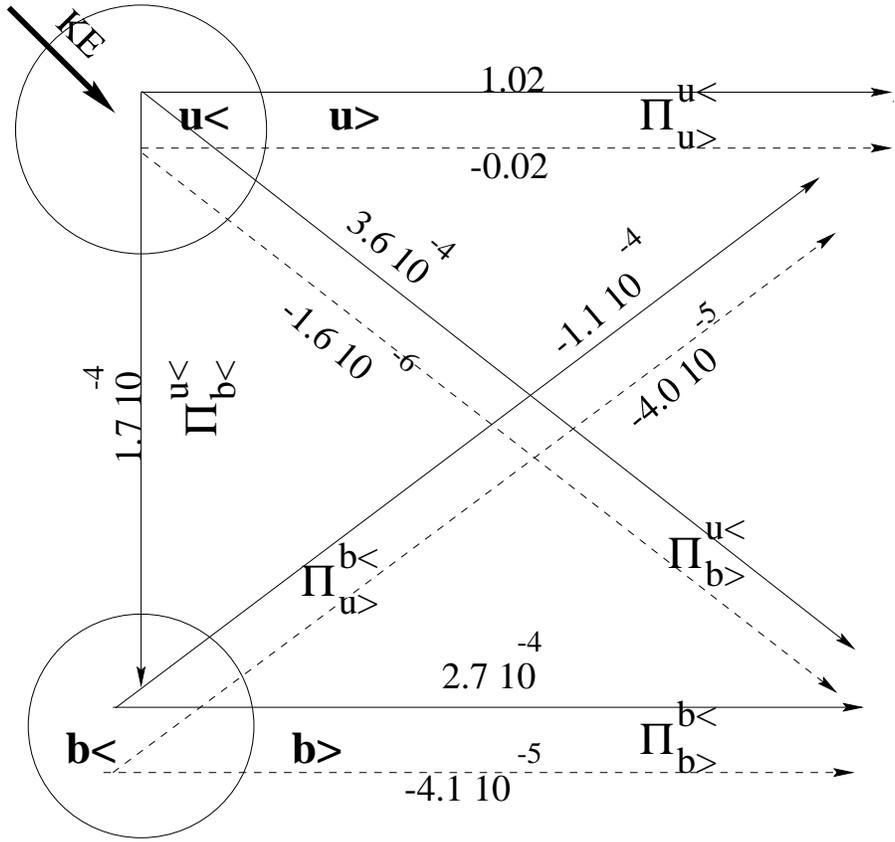,width=12cm,angle=0}}
        \vspace*{0.5cm}
\caption{Various energy fluxes in wavenumber space for parameter
$(r_A=5000,r_K=0.1,r_M=-0.1)$. The illustrated wavenumber spheres
contain ${\bf u<}$ and ${\bf b<}$ modes, while ${\bf u>}$ and ${\bf
b>}$ are modes outside these spheres.  The energy fluxes are from
large-scale $u$ to small-scale $u$ ($\Pi^{u<}_{u>}$), large-scale $u$
to small-scale $b$ ($\Pi^{u<}_{b>}$), large-scale $b$ to small-scale
$u$ ($\Pi^{b<}_{u>}$), large-scale $b$ to small-scale $b$
($\Pi^{b<}_{b>}$), and large-scale $u$ to large-scale $b$
($\Pi^{u<}_{b<}$).  The velocity fields at large-scale are forced, and
the net input energy is 1 unit.  The solid and dashed lines represent
the nonhelical and helical contributions respectively.}
\label{fig:helical_flux}
\end{figure}

\end{document}